\documentclass[a4paper]{article}

\usepackage{icrc2013}
\usepackage[english]{babel}
\newcommand{\ele}{\ensuremath{e^{+}}}
\newcommand{\pos}{\ensuremath{e^{-}}}
\newcommand{\eleca}{\texttt{EleCa}}
\newcommand{\hermes}{\texttt{HERMES}}
\newcommand{\etal}{\MakeLowercase{\textit{et al. }}} % "et al."
\title{ EleCa: a Monte Carlo code for the propagation of extragalactic photons at ultra-high energy}

\shorttitle{M. Settimo \etal The EleCa code for the propagation of UHE photons}

\authors{
Mariangela Settimo$^{1,2}$,
Manlio De Domenico$^{3,4,5}$
}

\afiliations{
$^1$ Universit\"at Siegen, Germany \\
$^3$ Laboratorio sui Sistemi Complessi, Scuola Superiore di Catania, Via Valdisavoia 9, 95123 Catania, Italy \\
$^4$ Dipartimento di Fisica e Astronomia, Universit\'a degli Studi di Catania, and Istituto Nazionale di Fisica Nucleare, Sez. di Catania, Via S. Sofia 64, 95123 Catania, Italy\\
\scriptsize{
$^{2}$ now at:  Laboratoire de Physique Nucl\'eaire et de Hautes Energies (LPNHE), Universit\'es Paris 6 et Paris 7, CNRS-IN2P3, Paris, France. \\
$^{5}$ now at:  Universitat Rovira i Virgili, Tarragona, Spain \\
}
}

\email{mariangela.settimo@gmail.com}

\abstract{
Ultra high energy photons play an important role as an independent probe of the photo-pion production mechanism by UHE cosmic rays. Their observation, or non-observation, may constrain astrophysical scenarios for the origin of UHECRs and help to understand the nature of the flux suppression observed by several experiments at energies above 10$^{19.5}$~eV. Whereas the interaction length of UHE photons above 10$^{17}$~eV is only of a few hundred kpc up to tenths of Mpc, photons can interact with the extragalactic background radiation leading to the development of electromagnetic cascades which affect the fluxes of photons observed at Earth. The interpretation of the current experimental results rely on the simulations of the UHE photon propagation.   
In this contribution, we present the novel Monte Carlo code ``EleCa'' to simulate the \emph{Ele}ctromagnetic \emph{Ca}scading initiated by high-energy photons and electrons. The distance within which we expect to observe UHE photons is discussed and the flux of GZK photons at Earth is investigated in several astrophysical scenarios.
}


\begin{document}
\maketitle

%Begin a section.
\section{Introduction}
Ultra-high energy (UHE) photons are expected to be produced in the interactions of UHE cosmic rays with matter, with the extragalactic background or in top-down models for the acceleration of UHE cosmic rays. For instance, one well known process producing UHE photons is the photo-pion production by UHE nuclei interacting with relic EBR photons (namely Greisen-Zatsepin-Kuz'min, GZK, effect~\cite{Greisen:1966jv,Zatsepin:1966jv}). 
Recent measurements of the UHECR energy spectrum~\cite{Zhang:2008zze,Settimo2012} have confirmed a flux suppression above 50 EeV (1\,EeV\,=\,$10^{18}$~eV) compatible with the expectation from GZK effect, even if other alternatives (e.g., a physical limit to the maximum acceleration at the source) can not be ruled out. Hence, the observation of UHE photons would be an independent evidence for the existence of the GZK effect, providing hints on the nature of UHECR, on astrophysical sources and on environmental conditions. 

No UHE photons have been identified so far and only upper limits to their flux have been placed~\cite{Settimo:2011zz,Rubtsov:2011zz} partly disfavoring exotic models for the origin of UHE cosmic rays. 

We present here a novel Monte Carlo simulation code, named \eleca (\emph{Ele}ctromagnetic \emph{Ca}cascading), to reproduce the propagation of extragalactic photons and their cascade development. A few codes for the propagation of cosmic rays and photons have been developed in the past, although some of them are still not publicly available. Most of these simulators are based on the solution of transport equations or on semi-analytical approaches. \eleca$\,$ is a \emph{stand-alone} code written in C++ with a modular structure which simplifies the interface with existing codes developed to simulate the propagation of UHECR (as for example~\cite{Armengaud:2006fx,hermes_epjp}).

In this paper we describe the structure of \eleca, presenting its salient features. The comparison with other existing simulation codes is also briefly discussed. Interaction processes and their implementation in \eleca$\,$ are treated in Section~\ref{sect:physics}, including a comparison with results available in literature. Finally, we show in Section~\ref{sect:applications} the application of \eleca\, to some representative astrophysical scenarios. 

\section{EleCa: \emph{Ele}ctromagnetic \emph{Ca}scading}\label{sect:physics}

\eleca~is an highly modular C++ code easy to interface with existing codes for the propagation of UHE nuclei. It simulates the propagation $\gamma$, \ele and \pos, i.e., their interactions with relic photons and the subsequent electromagnetic cascading processes.  The dominant energy-loss processes for UHE photons are the Pair Production (PP) and the Double Pair Production (DPP), responsible for the production of secondary electrons and positron (hereafter generally referred as electrons). Electrons are cooled down via the Inverse Compton Scattering (ICS), $e\gamma_b\rightarrow e\gamma$, or absorbed through the Triplet Pair Production (TPP), $e\gamma_b\rightarrow e\ele\pos$. 

\begin{figure}[!t]
\hspace{-0.4cm}
\includegraphics[width=0.50\textwidth]{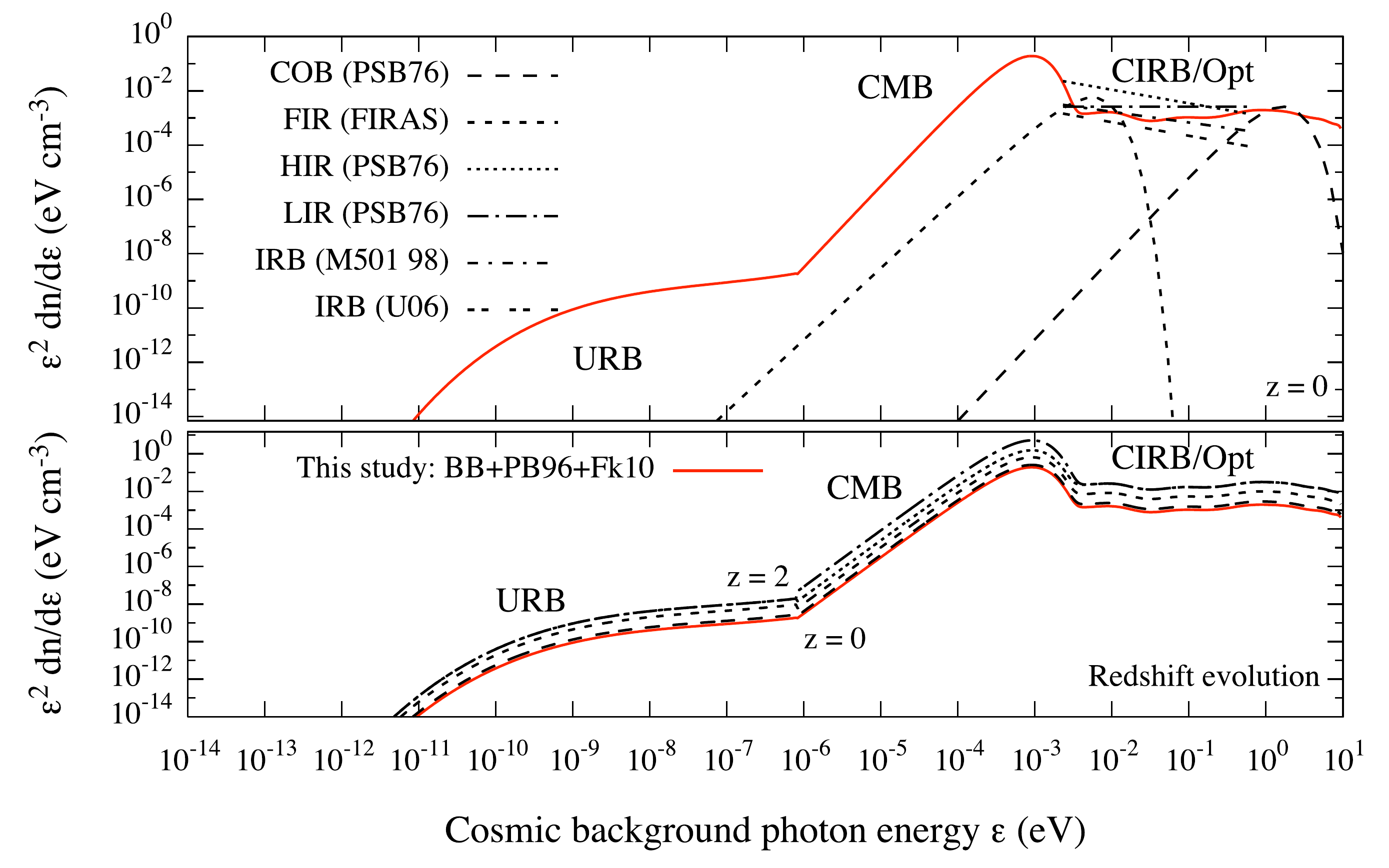}
\caption{Different parameterizations of extragalactic background radiation (EBR) as a function of relic photon energy. The red solid line indicates the EBR parameterization included in this study.}
\label{fig:EBR}
\end{figure}

Some models of EBR are shown in the top panel of Fig.\,\ref{fig:EBR}, as a function of the relic photon energy $\epsilon$ in the laboratory frame. The red solid line indicates the EBR spectrum adopted in this work. In particular, the blackbody model with temperature $T_{0}\simeq 2.725$~K is adopted for cosmic microwave background (CMB) while the semi-analytical ``model D'' in~\cite{finke2010modeling} is adopted for infrared/optical (CIOB). The model proposed in~\cite{protheroe1996new} is adopted for the universal radio background (URB).
For sake of completeness, we also show the models for COB (PSB76), for lower and higher infrared radiation (LIR and HIR, respectively) \cite{puget1976photonuclear}, and other infrared background models, derived from theoretical arguments or experimental observations \cite{epele1998propagation,funk1998upper,uryson2006ultra,fixsen2011probing}. The bottom panel of the same figure shows the evolution with redshift for different values of $z$, ranging from 0 to 2.

\begin{figure}[!t]
\begin{center}
\includegraphics[width=0.42\textwidth]{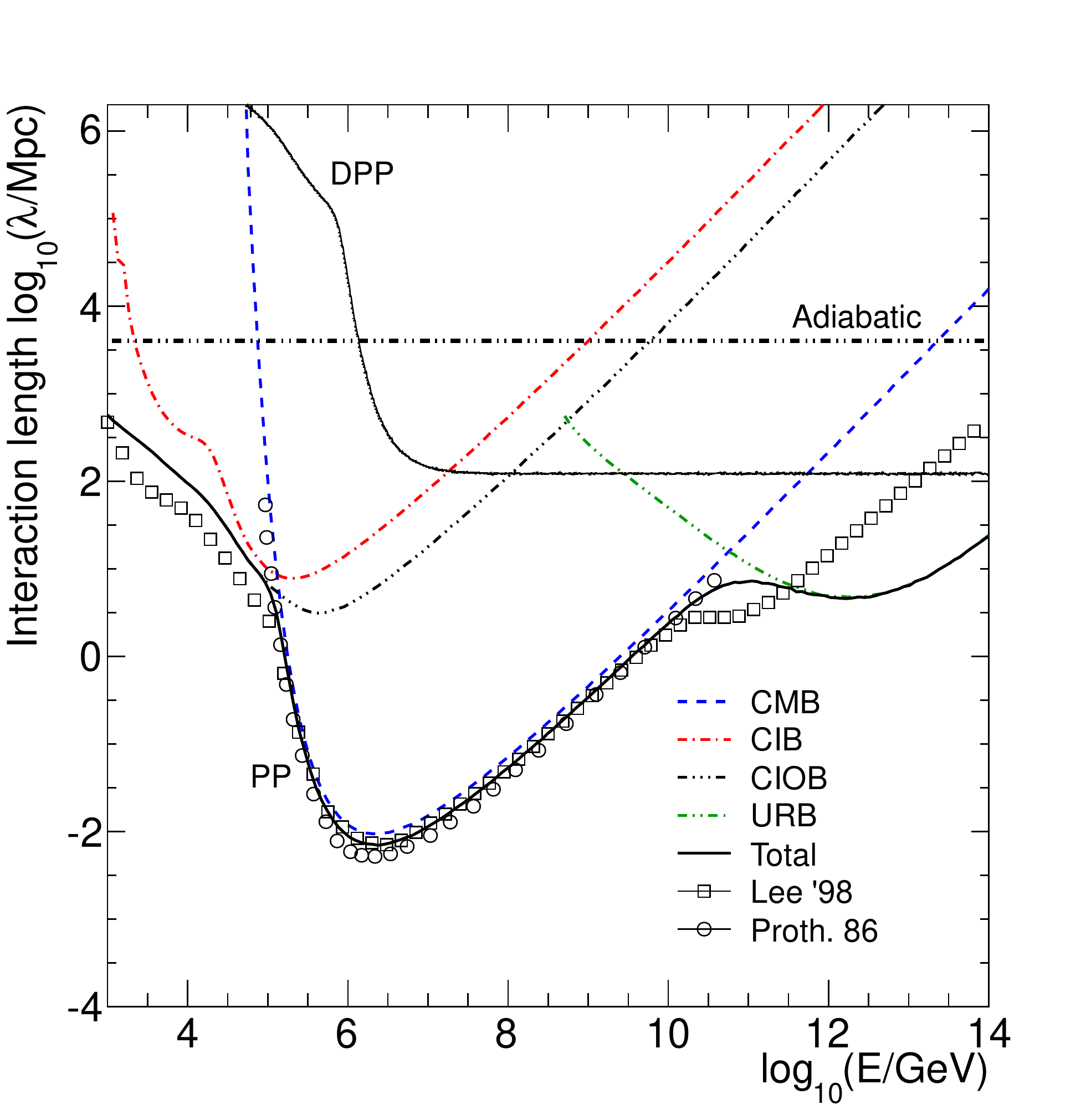}
\includegraphics[width=0.42\textwidth]{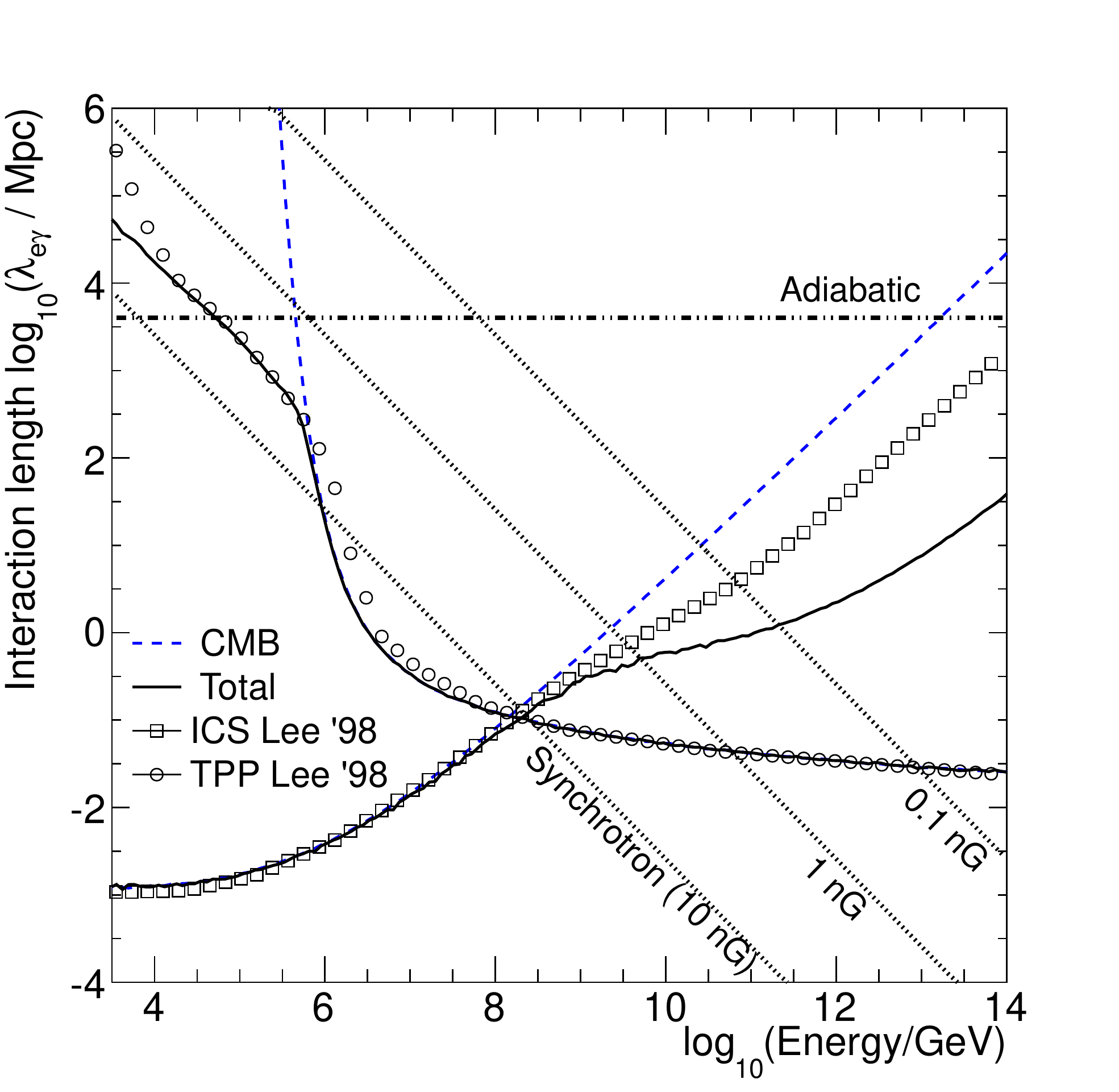}
\end{center}
\caption{Mean interaction lengths for photons (top) and electrons/positrons (bottom) for the whole EBR spectrum relevant for this study (solid). Dashed lines are given for each individual contributions to the EBR (see text).
The interaction length corresponding to the adiabatic and synchrotron energy losses are also shown as a dot-dashed and dotted lines respectively. }
\label{fig:lambda}
\end{figure}

The interaction length at the present epoch ($z=0$) for PP and DPP processes is shown in Fig.~\ref{fig:lambda} (top). The contribution of the whole EBR (solid line), as well as of each EBR component separately (dashed), is shown to emphasize the influence of each background radiation on the mean free path. As evident from the figure, photons with energy ranging below a few tens of GeV up to TeV, can propagate without interacting for distances of hundreds of Mpc, cooling down just because of adiabatic energy losses. On the other hand, the Universe becomes opaque to photons with energy of a few hundreds or thousands TeV, getting more transparent at energies in the EeV range, where the radio background is the main responsible for energy-loss processes. Previous results for PP~\cite{Lee:1996fp,1986MNRAS.221..769P} are also shown for comparison: the observed differences at very low and high energies are due to the different models assumed for the IRB and URB, respectively.

Similarly, the interaction lengths for ICS and TPP are shown in Fig.~\ref{fig:lambda} (bottom). The interaction lengths derived in previous works are also shown for comparison, leading to the same considerations already mentioned in the case of the PP and DPP. 
Energy losses due to the adiabatic expansion of the Universe and to average synchrotron radiation emissions \ele/\pos~\cite{Stanev:802298} are also shown in the same figure, for three different intensities of the magnetic field in the case of synchrotron.

Apart for these two energy losses, all the interactions, including DPP and TPP up to the highest energies, are treated as stochastic processes in \eleca, at variance with existing simulators publicly available to the community~\cite{Armengaud:2006fx, Kachelriess:2011bi}. 
The energy of primary particles and secondary products are estimated according to the cross section d$\sigma$/dE corresponding to each process.
Deflections due to the extragalactic magnetic field are taken into account by assuming the ``small angle'' approximation~\cite{Hooper:2006tn}, which is valid for magnetic fields with strength smaller than a few nG at the energy scale of interest in this paper.

\section{Photon fluxes at ultra-high energies}\label{sect:applications}
\subsection{Photon sources}\label{sect:phsource}

\begin{figure}[!t]
\begin{center}
\includegraphics[width=0.5\textwidth]{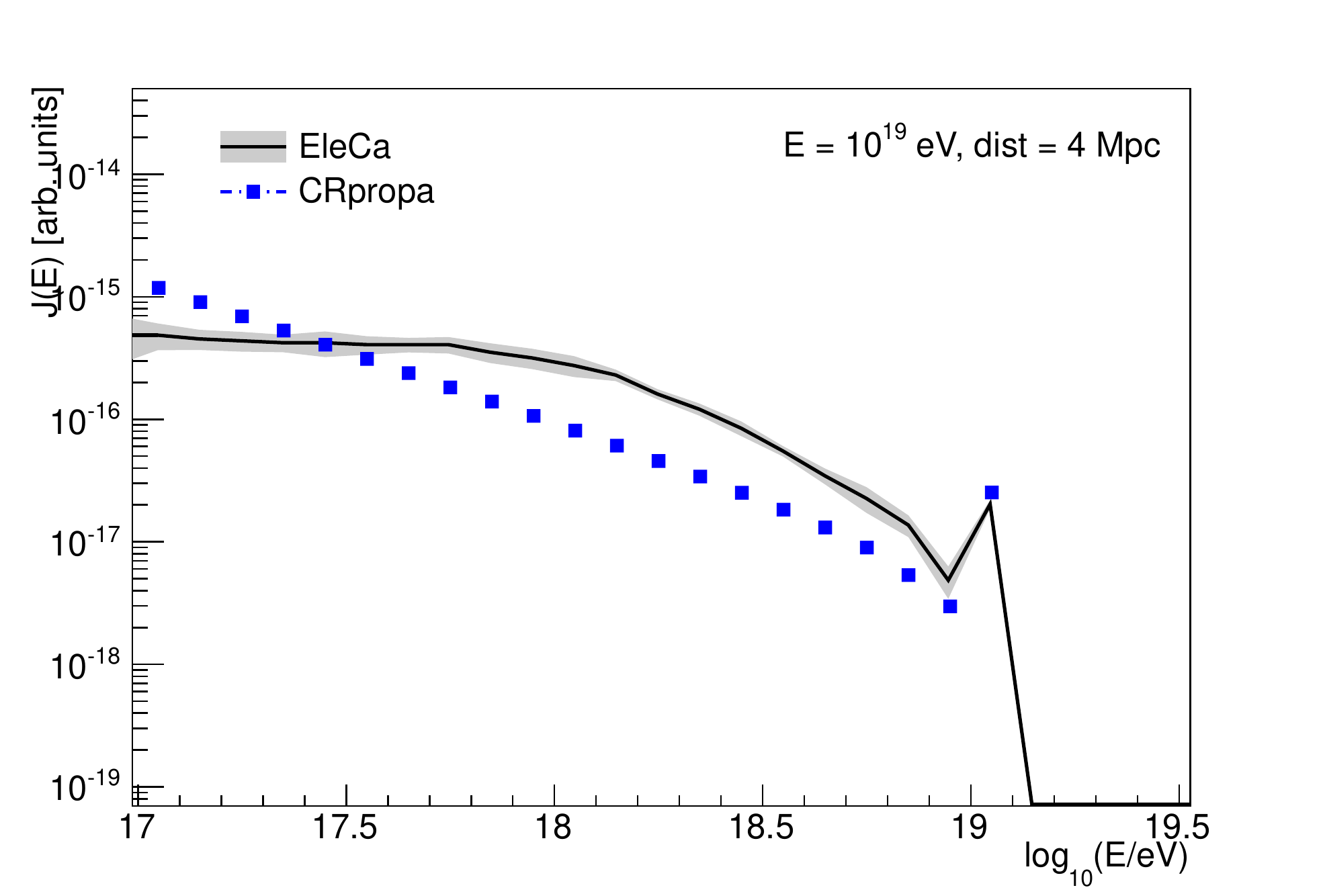}
\end{center}
\caption{Photon flux at Earth from a monochromatic photon source $E_{0}=10^{19}$~eV at 4~Mpc. CRpropa~\cite{crpropa2} results are shown for comparison as markers (see text).}
\label{fig:comparisonPhoton}
\end{figure}

We simulate the propagation of UHE photons injected from single monochromatic sources at different distances and we estimate the corresponding expected photon flux at Earth. An example is shown in Fig.~\ref{fig:comparisonPhoton} (lines) for the case of a source with $E_{0}=10^{19}$~eV at a distance of 4~Mpc. We perform 20 generations of the same sample of events from a monocromatic source. The average flux is shown as the solid line while the shaded area indicate the range of variability of the flux. This is the consequence of the stochastic nature of the Monte Carlo approach used in \eleca.  

For comparison, we also show the fluxes obtained from simulations with CRpropa (markers), assuming the same scenario. In this case, the predicted flux has not any variation between different generations, since the photon propagation is based on the numerical integration of transport equations~\cite{Lee:1996fp}, where photons are combined in predefined intervals of energy and distances according to their production sites. The observed differences between CRpropa and \eleca, may be additionally related to the different parameterization of IRB and URB, which are known to have a non-negligible impact on the propagation of electromagnetic particles. We refer to~\cite{crpropa2} for more details on the the modeling of background radiations. 
In fact, the discrepancies evident in Fig.~\ref{fig:comparisonPhoton} are compatible with the ones reported in Fig.~12 and Fig.~13 of~\cite{Lee:1996fp} when comparing the proposed code (later on implemented in CRpropa) and the Monte Carlo results in~\cite{Yoshida} and in~\cite{Protheroe96} respectively.  

\begin{figure}[!t]
\includegraphics[width=0.5\textwidth]{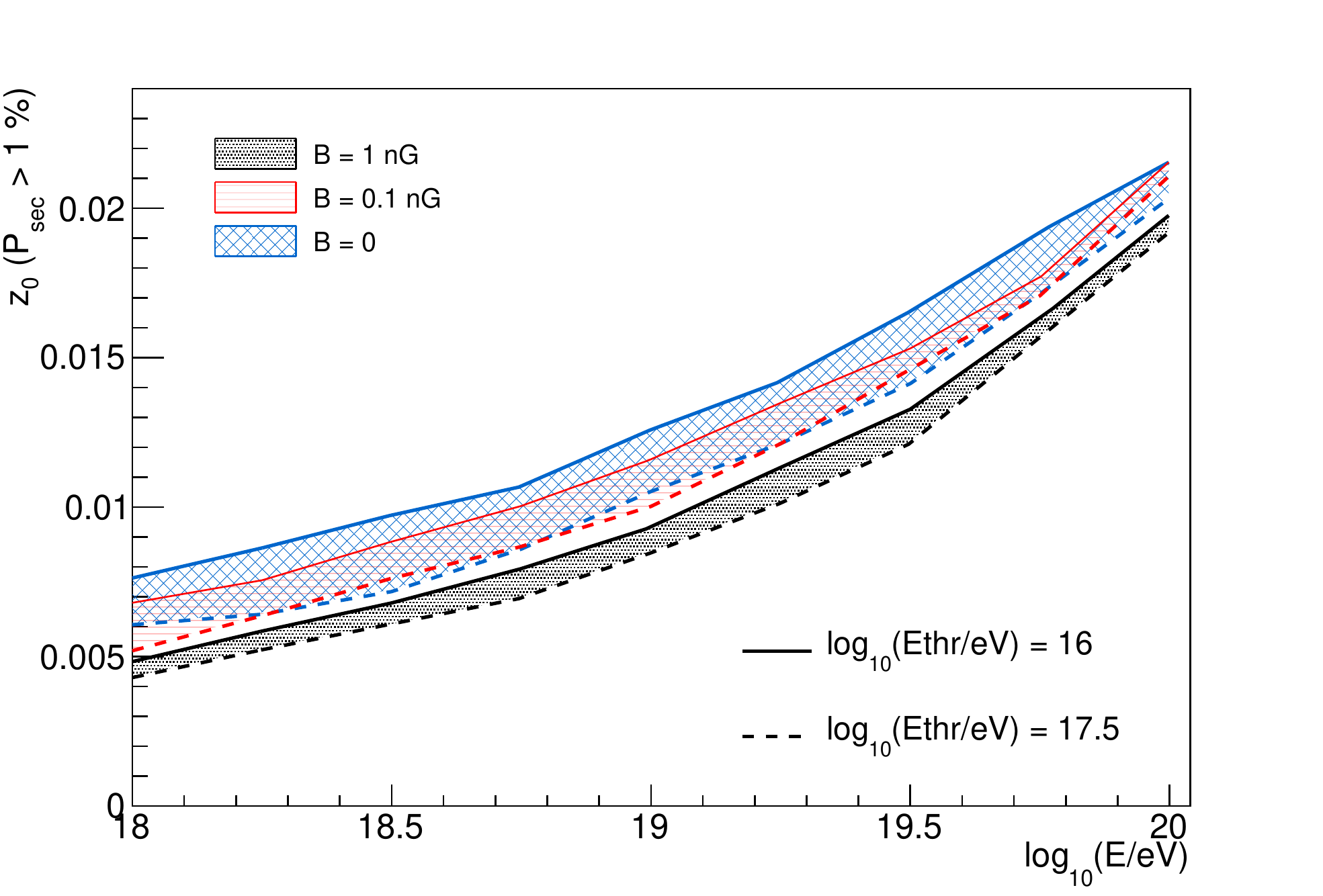}
\caption{Probability of observing at least 1 secondary $\gamma$ at Earth with energy larger than E$_{\rm{thr}}$ versus the energy (E$_{0}$) and the redshift (z$_{0}$) of the photon at the production site. See the text for further detail.}
\label{fig:horizon}
\end{figure}

In order to gain further insights into electromagnetic cascade propagation, we estimate the probability of observing a secondary photon at Earth versus the energy (E$_{0}$) and the redshift (z$_{0}$) of the photon at the production site. 
More specifically we define a probability $P_{\rm{sec}}$ to observe at least 1 $\gamma$ above the energy threshold E$_{\rm{thr}}$ for an initial configuration (E$_{0}$, z$_0$).  Our estimation of such a probability in some representative scenarios is shown in Fig.~\ref{fig:horizon}, for different intensity of intervening extragalactic magnetic field and for different values of E$_{\rm{thr}}$.
More specifically, we show the distance $z_{\rm{max}}$ corresponding to P$_{\rm{sec}} > $ 1\% versus the energy E$_{0}$. Different magnetic fields configurations are shown as bands, where each band is delimited by two example energy thresholds (10$^{17.5}$ and 10$^{16}$~eV). From these results, it is possible to define a maximum distance $z_{max}$ beyond which photon propagation does not significantly affect the photon flux at the Earth. A parameterization of $z_{\rm{max}}$ for the most conservative case (B=0) can be used to speed up the simulations by neglecting the cascade development induced by photons beyond the maximum distance z$_{\rm{max}}$(E).

\subsection{GZK photon fluxes}\label{sect:GZKfluxes}
Now we focus on another relevant mechanism responsible for the production of UHE photons, i.e., the GZK effect. In fact, nuclei interacting with relic photons experience baryonic resonances which produce neutral mesons, mainly pions, that in turn quickly decay to two $\gamma$. It is worth remarking that the number of UHE photons produced during nuclei propagation, as well as their production sites and their expected energy spectrum at Earth depend on the assumptions about the sources. More specifically, the injection mechanism of nuclei is not known: the energy spectrum at source is assumed to follow a power law $E^{-\gamma}$ with a cut-off at E$_{\rm{max}}$ that can be chosen as: 
\begin{equation}
E^{-\gamma}\Theta(E - E_{\rm{max}})
\label{eq:1}
\end{equation}
or 
\begin{equation}
E^{-\gamma}e^{-E/E_{\rm{max}}}. 
\label{eq:2}
\end{equation}

Moreover, the chemical composition at the source, as well as the relative flux normalization for each primary type,  the true distribution of sources and their redshift evolution are still unknown. The existing models of EBR and extragalactic magnetic fields (r.m.s. strength and coherence length of the turbulent component) add additional astrophysical and cosmological observables to vary in simulations.

The experimental observation (or non observation) of a photon flux compatible with the fluxes predicted by means of detailed simulations can be used to get insights into the sources and the extragalactic magnetic fields. 

Here, we consider several possible scenarios, varying the source injection, evolution and maximum acceleration, as well as nuclei abundances and intervening EGMF. The nuclei propagation is performed with \hermes, a novel Monte Carlo simulator described in detail in~\cite{hermes_epjp,hermes_icrc} and the produced GZK photons are successively propagated with \eleca\footnote{It is worth remarking that \eleca\, is a stand-alone code that can be used to propagate photons regardless of the specific nuclei propagator.}

In particular, we consider sources homogeneously distributed up to $z_{max}=2$ and the two injection mechanisms in eq.~(\ref{eq:1}) and eq.~(\ref{eq:2}). Different nuclear species at the source, namely proton, helium, oxygen and iron, are also considered. In one case, the impact of source evolution with redshift on the flux is investigated assuming star formation rate evolution.

\begin{figure}[!t]
\hspace{-0.4cm}
\includegraphics[width=0.50\textwidth]{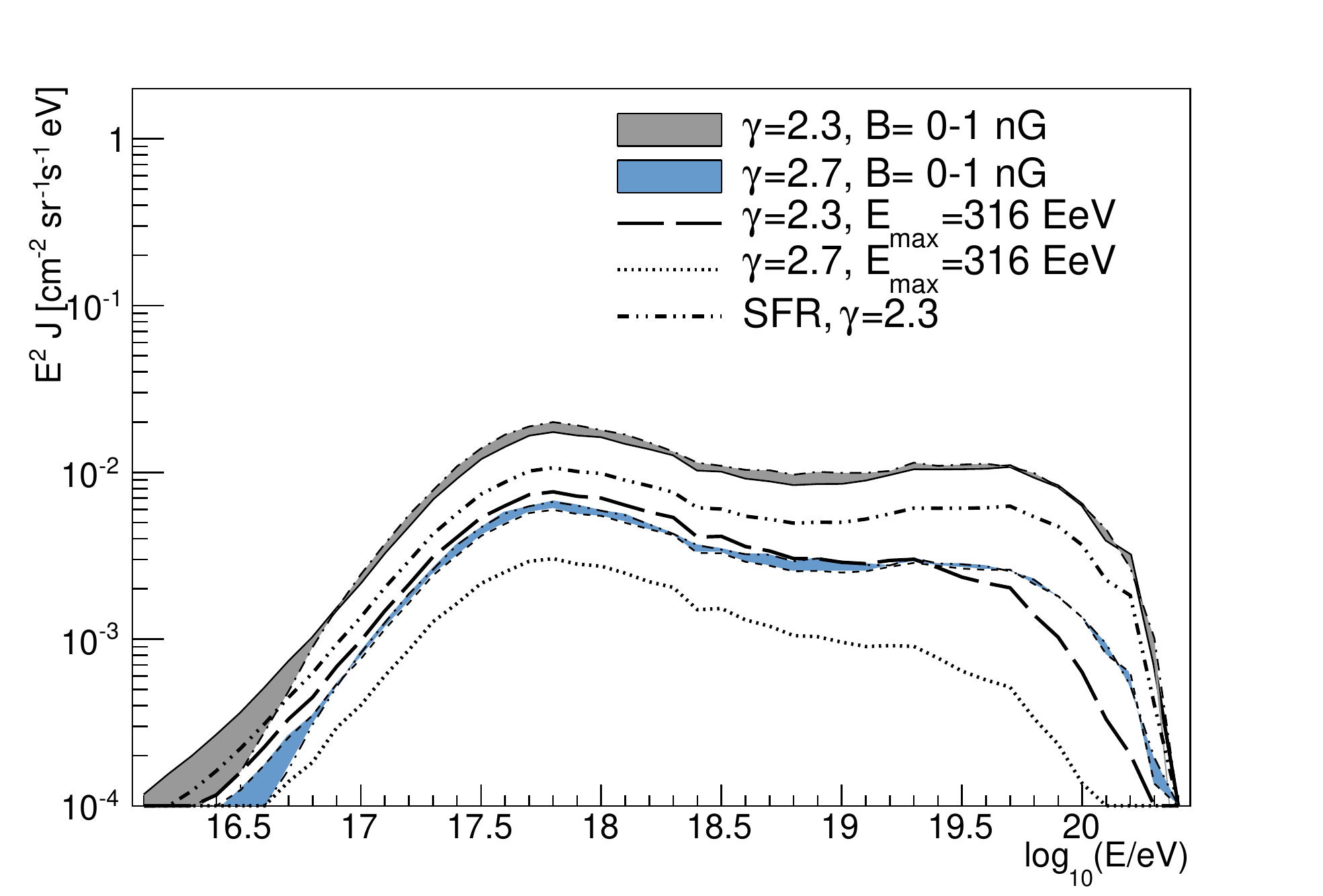}
\caption{Photon flux predicted at Earth assuming different scenarios at the source (i.e., varying spectral index, energy cut-off and redshift evolution) and intervening extragalactic magnetic fields. }
\label{fig:gzk_mixed}
\end{figure}

\begin{figure}[!t]
\hspace{-0.4cm}
\includegraphics[width=0.50\textwidth]{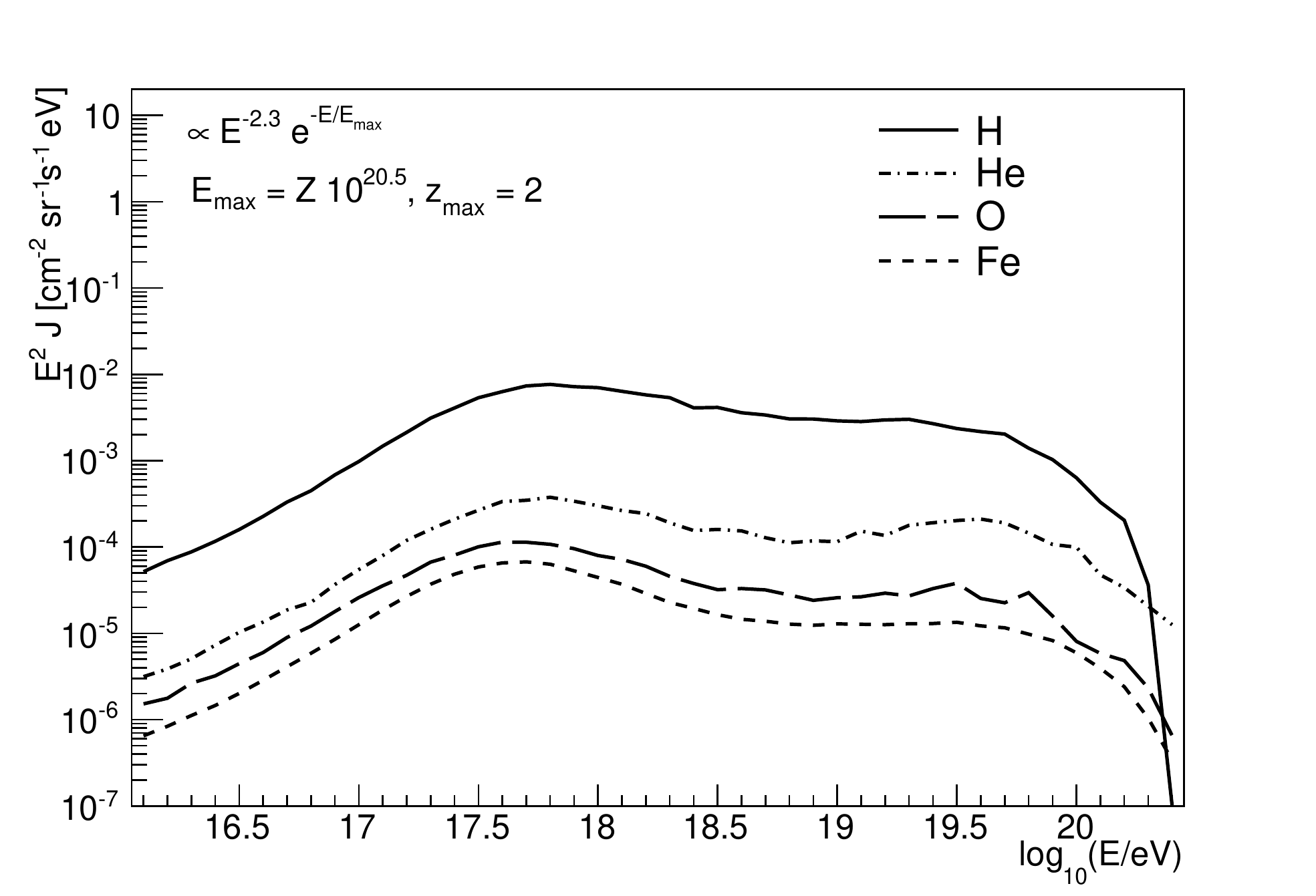}
\caption{Photon flux predicted at Earth injecting at the source different primary nuclear species.}
\label{fig:gzk_mass}
\end{figure}

If not specified otherwise, any other scenario does not account for source evolution, considering a sharp energy cut-off E$_{\rm{cut}} \propto 10^{21}$~Z~eV and a spectral index $\gamma = 2.3$. 
The fluxes for some representative simulated scenarios are shown in Fig.~\ref{fig:gzk_mixed} and Fig.~\ref{fig:gzk_mass}. The normalization is obtained by scaling the corresponding all-nuclei flux at 10$^{18.85}$~eV to the value measured by the Pierre Auger Observatory~\cite{Settimo2012}.

A change of the maximum energy correspondingly reduces the total flux of photons and suppresses the highest energies. A similar, although weaker, global effect is obtained in the case of steeper spectra. 
Moreover the impact of magnetic fields is also checked in fig.~\ref{fig:gzk_mixed} comparing the case of B=0 with the case of B=1 nG. It is worth mentioning here that the structured magnetic fields in the nearby Universe (within $\approx100$~Mpc) are not included in our simulations.  
A conclusive identification of optimal astrophysical scenarios to describe the current observations is out of the scope of the present work and will be the subject of successive studies. 

\section{Summary}

We have presented \eleca, a new code for the simulation of the development of electromagnetic cascades initiated by extragalactic photons, or electrons, at UHE. 
It is based on a pure Monte Carlo approach and it includes the dominant processes for photons and electrons in the highest energy regime. In the current version, the synchrotron emission is treated as a continuos process and the effect of magnetic deflections is included by adopting the small angle approximation. However a 3D version of \eleca$\,$, with a more realistic treatment of extragalactic magnetic fields and an optimization for the propagation of photons at the TeV scale is in progress. 

We have shown the impact of the cascading development on the detectability of photons coming from a source, or generally from any type of production sites.
In particular the flux of GZK photons at Earth can be predicted when \eleca$\,$ is combined with a nuclei propagator. We have shown here some examples of GZK photon fluxes derived for various astrophysical scenarios.

\vspace*{0.5cm}

\footnotesize{{\bf Acknowledgment:\\}{
The authors are grateful to H. Lyberis and M. Risse for useful discussions.\\
M.S. acknowledges support by the BMBF Verbundforschung Astroteilchenphysik and by the Helmholtz Alliance for Astroparticle Physics (HAP). 
}}

\end{document}